\documentclass[twocolumn,prb]{revtex4}

\usepackage{graphicx}
\usepackage{dcolumn}
\usepackage{bm}

\begin{document}

\title{
Comparative study of macroscopic quantum tunneling in Bi$_2$Sr$_2$CaCu$_2$O$_y$ intrinsic Josephson junctions with different device structures 
}

\author{K. Ota$^1$, K. Hamada$^2$, R. Takemura$^2$, M. Ohmaki$^2$, 
T. Machi$^3$, K. Tanabe$^3$, M. Suzuki$^{2,4}$, A. Maeda$^1$,}
\author{H. Kitano$^5$}
\email{hkitano@phys.aoyama.ac.jp}
 \affiliation{
 $^1$Department of Basic Science, University of Tokyo, Tokyo 153-8902, Japan\\ 
 $^2$Department of Electronic Science and Engineering, Kyoto University, Kyoto 615-8510, Japan\\
 $^3$Superconductivity Research Laboratory, ISTEC, Tokyo 135-0062, Japan\\
 $^4$Photonics and Electronic Science and Engineering Center, Kyoto University, Kyoto 615-8510, Japan\\
 $^5$Department of Physics and Mathematics, Aoyama Gakuin University, Kanagawa 229-8558, Japan\\
 }

\date{\today}

\begin{abstract}
We investigated macroscopic quantum tunneling (MQT) of 
Bi$_2$Sr$_2$CaCu$_2$O$_y$ intrinsic Josephson junctions (IJJs) 
for two device structures. 
One is a small mesa, which is a few nanometers thick with only two or three IJJs, and the other is a stack of a few hundred IJJs in a narrow bridge structure. 
The experimental results regarding the switching current distribution for the first switch from the zero-voltage state were in good agreement with the conventional theory for a single Josephson junction, indicating that the crossover temperature from thermal activation to the MQT regime for the former device structure was similar to that for the latter device structure. 
Together with the observation of multiphoton transitions between quantized energy levels in the MQT regime, these results strongly suggest that 
the observed MQT behavior is intrinsic to a single IJJ in high-$T_c$ cuprates and is independent of the device structure. 
The switching current distribution for the second switch from the first resistive state, which was carefully distinguished from the first switch, was also compared with respect to the two device structures. 
In spite of the differences between the heat transfer environments, 
the second switch exhibited 
a similar temperature-independent behavior for both devices up to a much higher temperature than the crossover temperature for the first switch. 
We argue that this cannot be explained in terms of self-heating caused by dissipative currents after the first switch. 
As possible candidates for this phenomenon, the MQT process for the second switch and the effective increase of the electronic temperature due to the quasiparticle injection are discussed.
\end{abstract}



\maketitle

\section{\label{sec:Intro}Introduction}
Macroscopic quantum tunneling (MQT) in a current-biased Josephson junction is a key phenomenon exhibiting quantum mechanical behavior in systems of macroscopic degrees of freedom. \cite{Leggett,Voss,Martinis1985} 
MQT is observed only at sufficiently low temperatures 
where thermal activation (TA) over the potential barrier is suppressed, 
while TA behavior is exhibited at temperatures higher than the relevant crossover temperature $T_{\rm cr}$. 
Recently, the observation of MQT has attracted great interest as a fundamental measurement for manipulating the so-called ``phase qubit", which is regarded as one of the promising candidates for future quantum information processing. \cite{Martinis} 
In such applications, some conventional superconductors (typically, Al and Nb) have been used for the fabrication of Josephson junctions. 
Thus, most experiments regarding the observation of MQT and the operation of phase qubits were performed only at sufficiently low temperatures ($\sim$ 0.1~K). 

The discovery of high temperature cuprate superconductors has drastically changed the situation due to the high critical temperature $T_c$ and the high critical current $I_c$ of these superconductors, which are one or two orders of magnitude higher than those of conventional superconductors. 
Particularly, intrinsic Josephson junctions (IJJs) in Bi$_2$Sr$_2$CaCu$_2$O$_y$ (BSCCO) single crystals have attracted great interest since a crystalline stack of the superconducting CuO$_2$ double layer and the insulating Bi$_2$O$_2$ (or SrO) layer forms a perfectly smooth tunnel junction on an atomic scale. \cite{Kleiner}
Indeed, recent experiments using a small stack of IJJs in a BSCCO single crystal have clearly indicated that MQT behavior can be observed up to nearly 1~K. 
\cite{Inomata,Jin} 

However, the use of stacked IJJs with almost identical $I_c$ gives rise to a serious problem where it is impossible to ensure that 
a switch to the first resistive state always occurs at the same junction when the bias current is repeatedly increased to obtain the probability distribution of the switching current. 
In addition, the anomalous enhancement of the MQT rate for 
``uniform-stack switching", where several tens of junctions in IJJs switch to the resistive states at almost the same time, remains unresolved. \cite{Jin}
Very recently, to avoid this complication of using stacked IJJs, 
a surface IJJ formed on the top of a small mesa with five IJJs was proposed by Li {\it et al}, \cite{Li}
since $I_c$ of the surface was significantly smaller than that of inner IJJs in the stack. 
The results from their experiments regarding the surface IJJ, which indicate that MQT behavior is observable only below $\sim$ 0.1~K, seem to point to another problem: the properties of the surface IJJ might be strongly influenced by a thick normal-metal electrode which is in direct contact with the surface IJJ. 
Thus, previous studies \cite{Inomata,Jin,Li} of MQT in stacked IJJs and surface IJJs have suggested that experiments involving the MQT phenomenon in IJJs should be performed more carefully by considering various influences originating from the device structure and the fabrication method of IJJs. 

In this paper, 
we present a comparative study of MQT in the following two types of IJJs. 
One is a small mesa, a few nanometers thick with only two or three IJJs, and the other one is a stack of a few hundred IJJs in a narrow bridge structure. 
The probability distribution of the switch from the zero-voltage state to the first resistive state in the multiple-branched current-voltage characteristics was carefully measured for both types of IJJs, confirming that the same junction switched to the resistive state in each event. 
We demonstrate that the MQT behavior for the surface IJJ on the former mesa structure survives up to around 0.4~K, which is similar to $T_{\rm cr}$ for the stacked IJJs in the latter bridge structure, in contrast to the results obtained in the study performed by Li {\it et al}. \cite{Li}
We also observed microwave-induced multiphoton transitions between quantized energy levels in the first switch for both types of IJJs, which were expected to appear in the MQT regime. 
The observation of both MQT and energy-level quantization (ELQ) in the first switch strongly suggests that these phenomena are intrinsic to a single IJJ of BSCCO and are independent of the device structure. 

Moreover, we performed similar switching-current distribution measurements for the second switch from the first resistive state to the second resistive state in the current-voltage characteristics. 
In spite of the differences in the heat transfer environments, 
the second switch for both devices showed a temperature-independent behavior, which was similar to MQT in the first switch, up to much higher temperatures (approximately an order of magnitude higher) than $T_{\rm cr}$ for the first switch. 
We discuss this unusual behavior from various points of view. 
In contrast to the recent paper by Kashiwaya {\it et al.}, \cite{Kashiwaya} 
the experimental results suggest that this behavior cannot be explained in terms of self-heating caused by dissipative currents after the first switch, 
as will be discussed below.

This paper is organized as follows. 
In Sec.~\ref{sec:model}, we briefly describe the conventional model of the escape from the zero-voltage state, which is based on the resistively and capacitively shunted junction (RCSJ) model. 
In Sec.~\ref{sec:expmt}, the preparation of both types of IJJs and the method employed in the switching-current distribution measurements are described. 
Section~\ref{sec:results} contains all experimental results and a discussion. 
Finally, we conclude this work in Sec.~\ref{sec:conclusion}.

\section{\label{sec:model}Model}
The current-voltage characteristics of a current-biased single Josephson junction are successfully explained on the basis of the so-called RCSJ model, which was independently studied by Stewart and McCumber, respectively.\cite{SMc_model}
In this model, the phase difference, $\theta$, across the Josephson junction is described by the following equation of motion, \cite{Tinkham}
\begin{equation}
\frac{d^2\theta}{dt^2}+\frac{1}{Q_0}\frac{d\theta}{dt}+\sin\theta-\frac{I}{I_c}=0, 
\label{eq:RCSJ}
\end{equation}
\noindent
where $I_c$ is a critical current and $Q^{-1}_0$ is a damping parameter given by $Q_0=\omega_{p0}RC$, respectively, $R$ and $C$ are the shunt resistance and the shunt capacitance, respectively, and $\omega_{p0}$ is the so-called Josephson plasma frequency given by 
\begin{equation}
\omega_{p0}=\sqrt{\frac{2eI_c}{\hbar C}}.
\label{eq:omega_P}
\end{equation}
\noindent
As is well known, the phase dynamics described by Eq.(\ref{eq:RCSJ}) is equivalent to a damped motion of a particle with a mass of $(\hbar/2e)^2C$ in a tilted washboard potential, $U(\theta)$, which is given by  
\begin{equation}
U(\theta)=-E_J(\cos\theta+\gamma\theta), 
\label{eq:U_theta}
\end{equation}
\noindent
where $E_J(=\hbar I_c/2e)$ is the Josephson coupling energy, and $\gamma=I/I_c$. 

In the underdamped region ($Q_0>1$), where the current-voltage characteristics become hysteretic, the zero-voltage state corresponds to the trapped particle at the bottom of a potential well (that is, $\theta$ is fixed at a certain value). 
When the washboard potential is tilted with increasing the bias current, 
the particle escapes from the well and starts rolling downhill 
(that is, $\theta$ changes at a rate $d\theta/dt=2eV/\hbar$). 
The bias current at which the junction switches from the zero-voltage state to the finite voltage state is referred to as switching current, $I_{\rm sw}$. 

In the absence of thermal and quantum fluctuations, such a switch always occurs at $\gamma=1$. 
However, when thermal (or quantum) fluctuations cannot be neglected, the switch is induced by a TA process over the potential barrier (or an MQT process through the potential barrier) for a smaller bias current $\gamma<1$. 
The potential barrier is given by \cite{Fulton}
\begin{equation}
\Delta U(\gamma)=E_J[\gamma(2\sin^{-1}\gamma-\pi)+2\cos(\sin^{-1}\gamma)]. 
\label{eq:barrier}
\end{equation}
\noindent
The escape rates induced by TA and MQT processes are given by 
\begin{eqnarray}
\frac{1}{\tau_{\rm TA}}&=&a_t\frac{\omega_p}{2\pi}\exp\Big(-\frac{\Delta U}{k_BT}\Big)
\noindent
and
\label{eq:tau_TA}\\
\frac{1}{\tau_{\rm MQT}}&=&\sqrt{120\pi\frac{36\Delta U}{5\hbar\omega_p}}\Big(\frac{\omega_p}{2\pi}\Big)\exp\Big(-\frac{36\Delta U}{5\hbar\omega_p}\Big),
\label{eq:tau_MQT}
\end{eqnarray}
\noindent
respectively\cite{Martinis1985}, where $\omega_p$ is the current-dependent Josephson plasma frequency,

\begin{equation}
\omega_p=\omega_{p0}(1-\gamma^2)^{1/4}.
\label{eq:omega_p}\\
\end{equation}
\noindent
$a_t(=4/[\sqrt{1+(Qk_BT/1.8\Delta U)}+1]^2)$ is a thermal prefactor, where $Q$ (=$Q_0(1-\gamma^2)^{1/4}$) is the current-dependent damping factor.
Eq.(\ref{eq:tau_MQT}) describes the escape rate induced by MQT without dissipation.
At high temperatures, the TA process is dominant, while a crossover to the MQT process occurs at the crossover temperature, which is approximately given by
\begin{equation}
T_{\rm cr}\sim\frac{\hbar\omega_p}{2\pi k_B}.
\label{eq:T_{cr}}
\end{equation}
\noindent
Experimentally, as a function of the bias current, the escape rate, $\tau(I)^{-1}$, is related to the probability distribution, $P(I)$, of the switching current through the following equation \cite{Martinis1985,Fulton}
\begin{equation}
\tau(I)^{-1}=\frac{1}{\Delta I}\frac{dI}{dt}\ln\frac{\sum_{i\geq I}{P(i)}}{\sum_{i\geq I+\Delta I}{P(i)}}, 
\label{eq:tau_PI}
\end{equation}
\noindent
where $\Delta I$ is a small current interval, the probability for the switch between $I$ and $I+\Delta I$ is given by $P(I)\Delta I$, and $dI/dt$ is the ramp rate of the bias current. 
Thus, it is expected that the standard deviation of $P(I)$, which is given by $\sigma(\equiv\langle(I_{\rm sw}-\langle I_{\rm sw}\rangle)^2\rangle^{1/2})$, is proportional to $T^{2/3}$ in the TA regime since 
$\Delta U(\gamma) \approx E_J4\sqrt{2}/3(1-\gamma)^{3/2}$ 
for $\gamma$ close to 1, while $\sigma$ is independent of $T$ in the MQT regime. \cite{Voss}
The fitting results of $P(I)$ using $I_c$ and an effective temperature at the escape event, $T_{\rm esc}$, also suggest that $T_{\rm esc}$ agrees with the bath temperature, $T_{\rm bath}$, in the TA regime, while $T_{\rm esc}$ becomes independent of $T_{\rm bath}$ in the MQT regime. \cite{Martinis1985}

When microwaves are irradiated onto the junction in the MQT regime, 
the distribution of $P(I)$ is divided into multiple peaks, which is attributed to the quantized energy levels in the potential well. 
The spacing between the $n$-th and the $n+1$-st energy levels, $E_{n,n+1}$, is obtained by solving the Schr\"{o}dinger equation under the boundary condition that 
the eigenfunction becomes zero when its energy is equal to the potential energy for a free-running side of the barrier. 
Numerical calculations suggest that $E_{01}$ is roughly given by 
$\hbar\omega_p$ as a function of $\Delta U(\gamma)$.\cite{Martinis1985}

\section{\label{sec:expmt}
Experimental method}
\subsection{\label{sec:3-1}Preparation of IJJ devices}
In this work, we studied two types of IJJs in BSCCO single crystals, which were grown by the floating zone (FZ) or the traveling-solvent floating zone (TSFZ) method \cite{Hanaguri,Watanabe} and annealed to be optimal doped. 

\begin{figure}
\includegraphics[width=\linewidth]{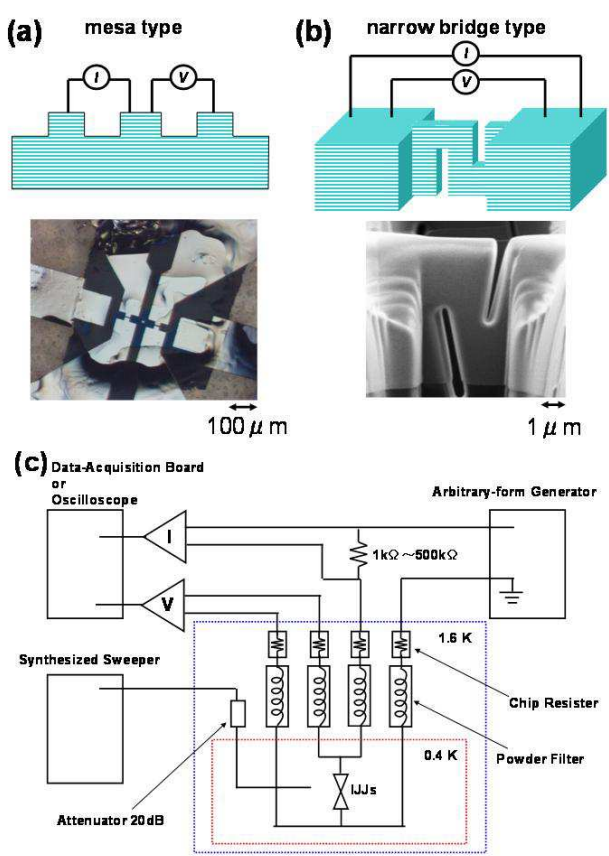}
\caption{(Color online). (a) (top) Schematic representation of the bias lead configuration and (bottom) a micrograph image of the mesa-type IJJ device. (b) (top) Schematic configuration and (bottom) a scanning ion micrograph image of the bridge-type IJJ device. (c) Schematic representation of the measurement setup.}
\label{fig:measurement}
\end{figure}

First, a small mesa, a few nanometers thick, with a few IJJs referred to as ``mesa-type IJJs" below, was fabricated as follows. 
A piece of the single crystal with a typical size of 0.5$\times$0.5$\times$0.01~mm$^3$ was glued to a sapphire substrate and cleaved in a vacuum chamber before evaporating a silver film to reduce the contact resistance. \cite{Suzuki1}
A small mesa structure with a height of a few nanometers was fabricated by using electron-beam (EB) lithography and Ar-ion milling techniques. 
The number of included junctions was confirmed by the observation of the $I-V$ characteristics, as shown below. 
Note that IJJs of this type are usually measured with a three-terminal configuration, as shown in Fig.~\ref{fig:measurement}(a). 

On the other hand, 
a few hundreds of stacked IJJs in a narrow bridge structure, which are referred to as ``bridge-type IJJs" below, was fabricated as follows. 
First, four gold electrodes were deposited on another piece of the BSCCO single crystal with dimensions of $1\times$0.1$\times$0.01~mm$^3$. 
Next, a narrow bridge $\sim$ 5~$\mu$m long and $\sim$ 1~$\mu$m wide was fabricated between two voltage terminals by using the focused ion beam (FIB) technique. 
Then, the sample was rotated by 90 degrees in order to etch the side wall of the bridge. 
Finally, two slits were fabricated at a distance of $\sim$ 1~$\mu$m by using FIB etching in order to separate a small stack of IJJs from the bridge structure. 
A typical scanning ion microscope image is shown in Fig.~1(b).\cite{FIB} 
The number of included junctions was estimated to be about 200 from the height of the stacks. 
Note that the bridge-type IJJs are measured with the usual four-terminal configuration, in contrast to the mesa-type IJJs. 
The details regarding the sample parameters are summarized in Table~\ref{tab:sample}. 

\begin{table*}
\caption{\label{tab:sample}
Various parameters of the fabricated IJJs. 
The lateral size and the number of IJJs for the mesa-type samples were determined by the mask size for EB lithography and the $I$-$V$ measurements, respectively, while those for the bridge-type samples were determined by the image of the scanning ion microscope. $T_c$ for each sample was determined by the zero-resistance, which was measured under a bias current far below the switching currents. $I_c$ (or $I_c^{2nd}$) was determined by the least square fitting of log($\tau^{-1}(I)$) for the first (or the second) switch (See also Ref.~[18]). $T_{\rm cr}$ was estimated from Eq.~(\ref{eq:T_{cr}}), and the uncertainty in $T_{\rm cr}$ was about 30\% due to the range variation of $\omega_p$, $\epsilon_r$, and $I_c$.}
\begin{ruledtabular}
\begin{tabular}{cccccccc}
Sample & Type & size ($\mu$m$^2$) & $N$ & $T_c$ (K) & $I_c$ ($\mu$A) & $T_{\rm cr} (K) $ & $I_c^{2nd}$ ($\mu$A) \\
\hline
M & mesa & 2$\times$2 & 2 & 74 & 10 & 0.4 & 41 \\
N & mesa & 2$\times$2 & 3 & 43 & 2.7 & 0.2 & -- \\
R & mesa & 2$\times$2 & 3 & 50 & 6.3 & 0.3 & -- \\
S23 & bridge & 1.1$\times$0.7 & $\sim$200 & 70 & 4 & 0.6 & -- \\
S301 & bridge & 1$\times$1 & $\sim$200 & 84 & 21.3 & 0.9 & 23.8 \\
S302 & bridge & 1$\times$1 & $\sim$200 & 84 & 12.1 & 0.8 & 22.3 \\
\end{tabular}
\end{ruledtabular}
\end{table*}

\subsection{\label{sec:3-2}Measurement technique and apparatus}
The probability distribution of the switching current from the zero-voltage state (or the first resistive state) was measured down to 0.4~K by using a $^3$He cryostat, as follows. \cite{Martinis1985}
First, a bias current is ramped linearly with time by using an arbitrary-waveform generator (Agilent 33220A) until the junction switches from the zero-voltage state to the first (or second) resistive state, and is reset to zero in order to wait for the next sweep. 
Next, as shown in Fig.~1(c), 
the ramped current and the voltage across the junction, which are amplified with a battery-powered preamplifier (Stanford Research Systems, SR560) or an operation amplifier (Analog Devices, AD524), are measured as a function of time by using a digital oscilloscope (Tektronics, TDS5054) or a data acquisition board (National Instruments, NI6251). 
The switching current, $I_{\rm sw}$, is recorded when the voltage is beyond a threshold value, which is typically set to 5~\% of the switching voltage in the first (or the second) switch above the zero-voltage state (or the first resistive state). 
Finally, the probability distribution of $I_{\rm sw}$ is obtained by repeating a large number of measurements of $I_{\rm sw}$ (typically 5000$\sim$10000 events). 

High-frequency noise through the four feed-lines connected to the IJJ sample is eliminated by using the following three stage low-pass filters. 
The first stage is a low-pass $\pi$-type feedthrough filter with a cutoff frequency of 1.9~MHz, which is operated at room temperature. 
We use a lossy stainless-steel coaxial cable (0.3~mm in diameter and 1~m in length) with the attenuation of 24~dB/m at frequencies above 1~GHz as the feed-lines to intervene between room temperature and 1.6~K, which is the base temperature of decompressed $^4$He. 
The second stage filter is a combination of a 1~k$\Omega$ chip resistor inserted at 1.6~K and the lossy coaxial cable, which was found to play a role similar to a $RC$ low-pass filter. 
The final stage filter is a home-made powder filter consisting of a 1.8~m long twisted pair of Manganin wires, which is coiled up in a 50~mm long copper tube filled with 100 mesh size SUS304 powder.
The room-temperature measurement performed by using a network analyzer (HP8753D) 
confirmed that the attenuation of this powder filter was more than 50~dB at frequencies above 1~GHz. 

The electromagnetic radiation from the environment is also carefully shielded as follows. 
First, the IJJ sample is mounted in a copper cell which is thermally anchored on the cold finger of a $^3$He pot. 
Next, both a $\mu$-metal and a permalloy shield are prepared in order to reduce stray magnetic fields. 
Moreover, all experimental components which do not need an AC power supply, such as the load resistor for monitoring the current, the battery-powered preamplifiers, the three stage filters, the sample cell, and the liquid-helium cryostat, are placed inside a shielded room (Japan Shield Co. Ltd., AR11), 
while the other instruments, which need an AC power supply, are placed outside the shielded room. 

Another coaxial transmission line consisting of several 0.085" coaxial cables and V-connectors was fed into the sample cell in order to observe the microwave irradiation effect up to 67~GHz. 
We also used a short superconducting NbTi coaxial cable for the thermal isolation of the sample cell and a coaxial attenuator (10~dB or 20~dB) for the reduction of thermal noise in the transmission line.

\section{\label{sec:results}Results and discussion}
\subsection{\label{sec:4-1}Properties of the first switch}
\begin{figure}
\includegraphics[width=\linewidth]{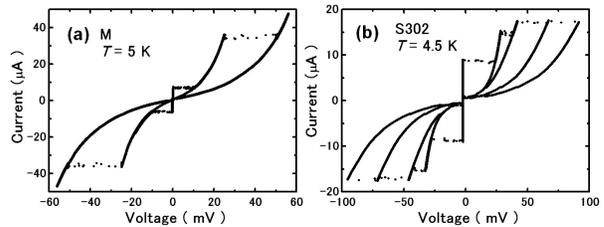}
\caption{$I-V$ characteristics of (a) sample M at 5~K and (b) sample S302 at 4.5~K.}
\label{fig:iv}
\end{figure}

Figures \ref{fig:iv}(a) and \ref{fig:iv}(b) show the typical $I-V$ characteristics of the mesa-type IJJs (sample M) and the bridge-type IJJs (sample S302), respectively. They were measured at a temperature of approximately 5~K. 
The negative voltage offset associated with the measurement electronics was subtracted in Figs. \ref{fig:iv}(a) and \ref{fig:iv}(b). 
As shown in Fig.~\ref{fig:iv}(a), only two resistive branches were observed in the $I-V$ characteristics for the mesa-type IJJs (sample M).
No more branches were observed up to higher bias (which is not shown here), which clearly indicates that the total number of junctions included in this sample was only two. 
We also note that the voltage jump ($\sim$10~mV) for the first switch in this sample is still larger than the previously reported value for similar mesa-type IJJs. \cite{Li} 

On the other hand, we found that the voltage jumps for the bridge-type IJJs 
were not always uniform, as shown in Fig.~\ref{fig:iv}(b). 
In the case of sample S302, the voltage spacing between the zero-voltage branch and the first resistive branch was slightly larger than the spacing between the second and third (or the third and forth) resistive branches. 
Since the latter voltage spacing (roughly 25~mV) is in good agreement with the previously reported values, \cite{Inomata,Jin} 
we consider that the former larger voltage spacing is attributed to the 
underdoping effect due to the residual Ga ions in the FIB etching process. 
A similar effect at the surface IJJs in the the mesa structure has been discussed by Zhao {\it et al}. \cite{Zhao} 
In contrast to the first voltage jump, the next voltage spacing between the first and second resistive branches was too narrow to be explained by the local overdoping. Rather, it is very similar to the behavior in a resistively shunted junction, implying that the junction which becomes resistive at the second switchings is locally shunted by a small resistance. 

We also found that the retrapping current (roughly, 3~$\mu$A) in the second branch was slightly larger than that in other branches. 
This feature is apparently similar to the behavior which was recently observed for the mesa-type IJJs with a small number of junctions (typically, less than 10). \cite{You,Takemura} 
Although You {\it et al.} \cite{You} suggested that this feature was attributed to the sheet critical current in a topmost layer of the mesa structure, 
we found that our result could not be explained by this scenario, since the observed difference of the retrapping currents between the first and the second branches was at least one order of magnitude smaller than the estimate of the sheet critical current (We roughly estimated it as 70~$\mu$A using the value given by You {\it et al.}). 
Rather, the larger retrapping current in the second resistive branch is qualitatively explained by the small shunted resistance, since the retrapping current is inversely proportional to the damping parameter, $Q_0(=\omega_{p0}RC)$. 
Thus, we consider that the unusual features observed for the second resistive branch are attributed to the small shunted resistance. 
It is also important to note that these features are not common to the other bridge-type samples. 
Although the origin of such a small shunted resistance is still unknown, it is clear that the small shunted resistance does not affect the switching current measurements. 

As shown in Figs.~\ref{fig:iv}(a) and \ref{fig:iv}(b), it was observed for both samples that the switching current for the first (or the second) switch was definitely smaller than that for the second (or the third) switch. 
The large difference between the first and the second (or the second and the third) switching currents allowed us to distinguish the corresponding switched junction from the rest of the junctions. 
We consider that the origin of the reduced switching current for the first switch is different in the mesa-type and the bridge-type samples.
Regarding the mesa-type samples, the first switch is considered to occur at a surface junction, as reported by Li {\it et al}.\cite{Li}
However, their interpretation that the surface layer was intentionally underdoped seems to be difficult to explain the observed small voltage jump. 
Rather, we consider that a kind of proximity effect plays an important role in reducing the switching current at the surface junction.\cite{Suzuki2}
On the other hand, in the bridge-type sample, the residual Ga ions seem to reduce the switching current in the first and the second switch.
Although the microscopic details of the influence exerted by such residual Ga ions are unclear, a plausible explanation is that the residual Ga ions make nearby junctions slightly underdoped, leading to the reduction of the critical currents.

\begin{figure}
\includegraphics[width=\linewidth]{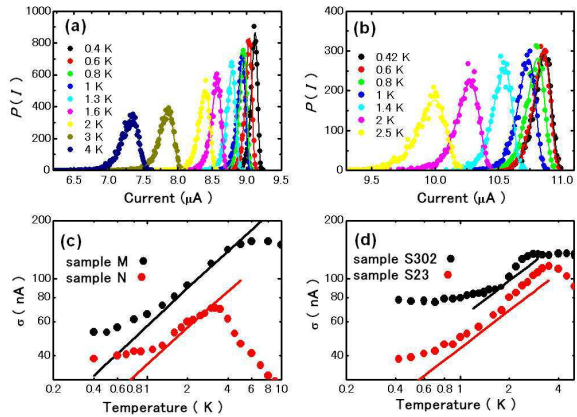}
\caption{(Color online). (a) Switching current distribution, $P(I)$, for the first switch of sample M. (b) Similar plots of $P(I)$ for the first switch of sample S302. (c) Solid symbols are the temperature dependence of the standard deviations, $\sigma(T)$, of $P(I)$ for the mesa-type IJJs (sample M and N), while  solid lines represent the calculation in the TA regime. (d) Similar plots of $\sigma(T)$ for the bridge-type IJJs (sample S302 and S23).}
\label{fig:sd1st}
\end{figure}

These results clearly suggest that the quality of the fabricated IJJ devices is sufficiently high for investigating the switching properties in the TA and MQT regimes. 
Based on these observations, the respective switching current distributions, $P(I)$, for the first switch in both samples (M and S302) were measured for temperatures down to 0.4~K, as shown in Figs.~\ref{fig:sd1st}(a) and \ref{fig:sd1st}(b). 
We found that the experimental data (solid symbols) for $P(I)$ below 4~K fitted the theoretical calculations (solid lines), where we used $I_c$ and $T_{\rm esc}$ as fitting parameters for the single junction model in the TA regime and assumed simply that $a_t$ was 1.\cite{fit}  
Figures \ref{fig:sd1st}(c) and \ref{fig:sd1st}(d) show the temperature dependence of the standard deviation of $P(I)$, $\sigma(T)$, for several samples. 
The experimental results (solid symbols) of $\sigma(T)$ were also compared with the calculations (solid lines) in the TA regime. 
We observed that the behaviors of $\sigma(T)$ for both mesa- and bridge-type IJJs were divided into the following three temperature regions; 
(i) the higher temperature region (above approximately 4~K), where $\sigma(T)$ increased as the temperature decreased, 
(ii) the intermediate temperature region (between approximately 4~K and 1~K), where $\sigma(T)$ decreased as the temperature decreased, 
(iii) the lower temperature region (below approximately 1~K), where $\sigma(T)$ began to saturate as the temperature decreased further. 

As shown in Fig.~\ref{fig:sd1st}(c), we found not only that $\sigma(T)$ for the mesa-type IJJs (sample M and N) was proportional to $T^{2/3}$ in the region (ii), but also that $\sigma(T)$ was in good agreement with the calculation in the TA regime, where it is assumed that $T_{\rm esc}=T$ as well as that $a_t=1$. 
This strongly suggests that the first switch for the mesa-type IJJs are successfully described by the conventional single junction model. 
The crossover temperature, $T_{\rm cr}$, was estimated from Eq.(\ref{eq:T_{cr}}) as 0.4~K and 0.2~K for sample M and N, respectively, as listed in Table~\ref{tab:sample}. 
We found that the estimated $T_{\rm cr}$ was slightly lower than the temperature below which $\sigma(T)$ showed a saturated behavior in the crossover region from the TA process to the MQT process. 
Unfortunately, the reason for this small discrepancy between the two temperatures remains unknown at present. 
However, our results clearly suggest that the crossover from the TA process to the MQT 
process occurs in region (iii).
Furthermore, the observed behavior in region (i) is qualitatively explained by the phase retrapping-diffusion process. \cite{Krasnov}

The results for $\sigma(T)$ for the bridge-type IJJs (sample S302 and S23) were found to be almost the same as those for the mesa-type IJJs, as shown in Fig.~\ref{fig:sd1st}(d). 
However, a small discrepancy in $\sigma(T)$ between the experimental data and the calculations in the TA regime was observed in region (ii). 
At present, the following three explanations can be considered for this discrepancy. 
One is the contribution of the thermal prefactor, $a_t$, which was neglected in the above analyses. 
Since $a_t$ is dependent on $Q$ \cite{Buttiker1983}, 
new fitting analyses including the contribution of $a_t$ suggest that $Q$ of the bridge-type IJJs was more than 250, while that of the mesa-type IJJs was $30\sim70$. 
As discussed below, $Q$ was estimated to be 100 for the bridge-type IJJs from the resonant peak under microwave irradiation.
This suggests that the above fitting analyses tend to overestimate the $Q$ value. 
The second one is the effect of large Josephson junctions, which becomes important when the junction size is larger than the Josephson penetration depth, $\lambda_J$. 
A previous study, in which a large mesa of IJJs (from 10$\times$10 to 40$\times$40~$\mu$m$^2$) was utilized, suggested that the switching distribution was further broadened by the large junction effects. \cite{Kitano} 
By using the fitting results in the TA regime and the assumed value of $\lambda_{ab}(\sim 250~{\rm nm})$, $\lambda_J$ was estimated to be about 0.4~$\mu$m for sample S302, which is indeed smaller than the junction size (1~$\mu$m). 
However, we found that the estimated $\lambda_J(\sim 0.8~\mu{\rm m})$ for sample M was also smaller than the junction size (2~$\mu$m), suggesting that this effect is unable to explain the small discrepancy in $\sigma(T)$ observed only in the bridge-type IJJs. 
As the third candidate, we found that the discrepancy in $\sigma(T)$ became more prominent when the switching current for the first switch was closer to that for the second switch. 
This implies that the coupling effect between adjacent layers, which is neglected in the conventional single junction model, might be closely related to the observed discrepancy in $\sigma(T)$. 

In order to confirm the realization of the MQT process more clearly, we measured $P(I)$ under microwave irradiation at the lowest temperature. 
In the MQT regime, where the energy levels in the potential well are quantized, the transition from the ground state to the first excited state can be induced by microwave radiation whose frequency is equal to the energy level spacing in the potential well. 
The numerical calculation of the Schr\"{o}dinger equation suggests that the energy level spacing, $E_{\rm 01}$, between the ground state and the first excited state is roughly given by $\hbar\omega_p$, as described in the previous section. 
Since $\hbar\omega_p/2\pi$ easily exceeds 100 GHz in the case of IJJs, it is difficult to observe the transition across $E_{\rm 01}$ due to the single photon absorption process. 
However, the multiphoton process based on using a microwave frequency equal to one for integer $E_{\rm 01}$ is expected to work well even in the present case. \cite{Wallraff} 

\begin{figure}
\includegraphics[width=\linewidth]{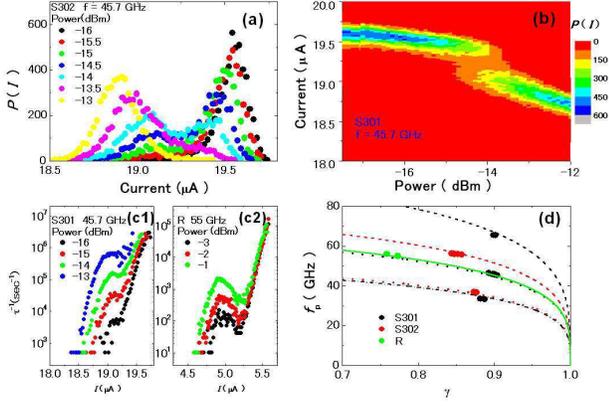}
\caption{(Color online). (a) $P(I)$ for the first switch of sample S301 under several different levels of microwave irradiation . (b) Equi-intensity contour in the switching current versus microwave power plane for three-photon absorption, measured at 45.7~GHz for sample S301. (c1)(c2) $1/\tau(I)$ under microwave irradiation for samples S301 and R, respectively. (d) Plots of the applied microwave radiation versus the normalized resonance current for several samples (S301, S302, and R). Solid, dashed, dotted and dash-dotted lines represent the calculations for single-photon, two-photon, three-photon and four-photon absorptions, respectively, which are based on Eq.~(\ref{eq:omega_p}).}
\label{fig:mw1st}
\end{figure}

Figure~\ref{fig:mw1st}(a) shows the measurements of $P(I)$ for sample S301 under microwave radiation with a frequency of 45.7~GHz. 
We observed that a resonant peak developed in the lower current region of $P(I)$ as the microwave power increased, which corresponded to the MQT process from the excited state. 
We estimated the current-dependent plasma frequency $\omega_p(=2\pi\times$136~GHz) at $I$=19.0~$\mu$A by using the values of the critical current $I_c$=21.3~$\mu$A and the junction capacitance $C$=39.9~fF, which is estimated from the junction size and the relative permittivity $\epsilon_r(\sim$5). \cite{yuudenritu} 
Note that the estimated $\omega_p$ is three times as large as the applied microwave frequency at 45.7~GHz, suggesting that a three-photon process was detected. 

Figure \ref{fig:mw1st}(b) shows an equi-intensity contour in the switching current versus the microwave power plane for three-photon absorption, which was measured at 45.7~GHz for sample S301. 
Here, the microwave power is measured at the output port of the microwave synthesized sweeper. 
Since the attenuation of the microwave power in the transmission line, including some connectors and an attenuator, depends strongly on the microwave frequency, it is difficult to compare the power at the sample position between the different frequencies.
We also plot the escape rate, $\tau(I)^{-1}$, which was obtained from $P(I)$ for sample S301 and sample R, as shown in Figs.~\ref{fig:mw1st}(c1) and \ref{fig:mw1st}(c2), respectively. 
Note that $\tau(I)^{-1}$ for sample S301 corresponds to a three-photon process ($\omega_p/2\pi$=45.7$\times$3~GHz), while that for sample R corresponds to a single-photon process ($\omega_p/2\pi$=55~GHz). 
We found that the range of $\tau(I)^{-1}$ for sample S301 slightly shifted toward the lower current region as the microwave power increased, in addition to the growth of the resonant peak due to the MQT process from the excited state. 
This is in sharp contrast to the results for $\tau(I)^{-1}$ for sample R, where no shift in the range of $\tau(I)^{-1}$ was observed as the microwave power increased. 
We found that these behaviors were related to the fact that the microwave AC current affects the switching characteristics of the junctions. \cite{Fistul,Guozhu} 
In this case, the switching current distribution is strongly affected by the microwave power. 

In order to take into account the influence of the ac current on the switching current, we corrected the apparent difference between the switching current under microwave irradiation and that without microwave irradiation. 
As shown in Fig.~\ref{fig:mw1st}(d), we plotted the resonant frequency, $f_p(\equiv\omega_p/2\pi)$, which was observed for several multiphoton processes as a function of the corrected bias current at which $\Delta\tau(I)^{-1}/\tau(I,0)^{-1}[\equiv \{\tau(I+I_{\rm shift},p_{\rm mw})^{-1}-\tau(I,0)^{-1}\}/\tau(I,0)^{-1}]$ showed a resonant peak.
Here, $I_{\rm shift}$ and $p_{\rm mw}$ indicate the apparent shift in the switching current which compensates the contribution of the ac current and the irradiated microwave power, respectively. 
The solid lines in Fig.~\ref{fig:mw1st}(d) represent the calculation based on Eq.~(\ref{eq:omega_p}), which is in good agreement with our experimental results. 
We also estimated the $Q$ value for such resonance peaks by fitting them with a Lorentzian, which suggested that $Q$ was 80$\sim$110 for sample S301 and 50$\sim$90 for sample S302, respectively. 
Regarding sample R, for lower $T_{\rm cr}(\sim$0.3~K), we observed only a broad step-like feature in $\Delta\tau(I)^{-1}/\tau(I,0)^{-1}$ for sample R even at the lowest temperature, in contrast to the results for samples S301 and S302 with higher $T_{\rm cr}$. 
Similar results have already been reported by Martinis {\it et al}, 
\cite{Martinis1985} 
suggesting that the measured temperature was not far below $T_{\rm cr}$. 

It has been highlighted recently that the microwave-induced multi-peaked switching distribution and the Rabi-type oscillation can be explained by using the classical model of Josephson junctions.\cite{Gronbech}
This scenario is likely to become more important for multilevel systems with larger dissipation. 
In order to distinguish between a two-level system and a multilevel system, it is useful to evaluate $\Delta U/\hbar \omega_p$.\cite{Lisenfeld} 
In our samples, $\Delta U/\hbar \omega_p$ was estimated to be about 5 for samples S301 and S302 and about 10 for sample R at the lowest temperature.
The values for the former samples, S301 and S302, are typical when observing Rabi oscillation in a two-level system, suggesting that our results shown in Figs.~\ref{fig:mw1st}(a) and \ref{fig:mw1st}(b) can be regarded as indicating quantized energy levels.
On the other hand, the value for sample R is closer to that for a multilevel system.
This behavior can be explained by considering the fact that $\Delta U/\hbar \omega_p$ decreased as the bias current increased.
Since the switching current decreased as the temperature increased near and above $T_{\rm cr}$, the larger value of $\Delta U/\hbar \omega_p$ for sample R suggests that the measured temperature is not far below $T_{\rm cr}$, as discussed above.
Thus, a clearer indication of ELQ for sample R is expected to be observed if the temperature is decreased further.
Based on these results, we concluded that the indication of MQT and ELQ in the first switch was essentially common to both device types, which suggests that it was intrinsic to a single IJJ of BSCCO and independent of the device structure.

\subsection{\label{sec:4-2}Properties of the second switch}
\begin{figure}
\includegraphics[width=\linewidth]{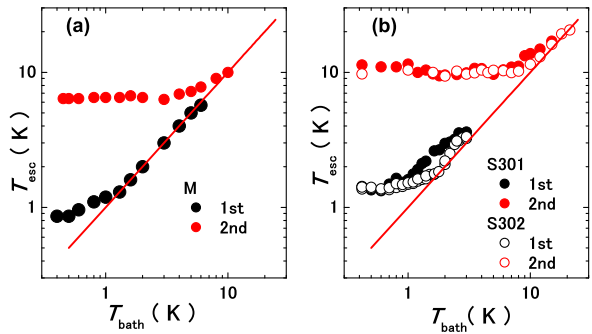}
\caption{(Color online). Plots of the effective escape temperature $T_{\rm esc}$ versus the bath temperature $T_{\rm bath}$ for both the first and the second switch of (a) the mesa-type IJJs and (b) the bridge-type IJJs, respectively. The solid lines represent $T_{\rm esc}=T_{\rm bath}$}
\label{fig:Tesc}
\end{figure}

We also measured the temperature dependence of $P(I)$ for the second switch from the first resistive state in the same way. 
Figure~\ref{fig:Tesc} shows the plots of $T_{\rm esc}$ versus the bath temperature, $T_{\rm bath}$, where $T_{\rm esc}$ was estimated from Eq.~(\ref{eq:tau_TA}). 
We found that the decrease of $T_{\rm esc}$ for the first switch saturated below a certain temperature, $T_{s}^{\rm 1st}$ (roughly $\sim$1~K), while that for the second switch saturated below a much higher temperature, $T_{s}^{\rm 2nd}$ (roughly $\sim$8~K). 
Surprisingly, these behaviors were found to be common to both types of IJJs. 
It should be noted that $T_{s}^{\rm 2nd}$ was also much higher than the estimated crossover temperature, $T_{\rm cr}^{\rm 2nd}$ ($\sim$0.8~K for the mesa-type samples and $\sim$1.3~K for the bridge-type samples), for the second switch. 
Note that the saturation at higher temperature is not attributed to the heating of the contact resistance, which was less than 100~$\Omega$, since its influence was not observed in the first switch.

Similar behaviors have recently been investigated by Kashiwaya {\it et al.} \cite{Kashiwaya}
They argued that this unusual behavior was attributable to the self-heating effect in the resistive state after the first switch. 
However, we found that this interpretation could not explain our results successfully since the self-heating effect was strongly dependent on the heat transfer environment in the device structure, which was significantly different in the mesa-type IJJs and the bridge-type IJJs. 
The problem of self-heating for the mesa structure has been numerically investigated by Krasnov {\it et al.} \cite{Krasnov2001} 
They estimated the magnitude of the self-heating by using the two-dimensional solution of a heat diffusion equation for a small circular mesa on top of a semi-infinite single crystal. 
They also suggested that the contribution from the metallic electrode on top of the mesa and that from an insulating layer separating the electrode from the bulk crystal played an important role as parallel heat channels. 
On the other hand, it seems that the heat diffusion process in the bridge-type IJJs with a few hundred junctions is quite different from that in the mesa-type IJJs. 
\cite{Kashiwaya_private}
For instance, the heat diffusion process along the stacked direction (that is, along the $c$-axis of the crystal), becomes more important since the stack of IJJs is far away from the remaining part of the bulk crystal. 
In addition, there is no effective parallel heat channel since the stack of IJJs is surrounded by vacuum and is located at a sufficient distance from the electrode. 
This strongly suggests that the magnitude of the self-heating in the bridge-type IJJs should be much greater than that in the mesa-type IJJs. 

If we assumed that the thermal resistance, $R_{\rm th}$, was independent of the dissipated power, $P_{\rm dis}$, we can estimate the temperature rise, $\Delta T_{\rm h}(=R_{\rm th}\times P_{\rm dis})$, at a current of the second switch using the recent experiments of the self-heating which were measured at a larger dissipated power. \cite{Verreet,Wang}
For instance, Verreet {\it et al.}\cite{Verreet} estimated that $R_{\rm th}\sim$160~K/mW at $P_{\rm dis}\sim$150~$\mu$W for the mesa-type IJJs. 
This suggests that $\Delta T_{\rm h}$ for sample M is $\sim$0.14~K for $P_{\rm dis}=$0.9~$\mu$W in our measurements at the lowest temperature. 
On the other hand, Wang {\it et al.} \cite{Wang} estimated that $R_{\rm th}\sim$450~K/mW at $P_{\rm dis}\sim$70~$\mu$W for the bridge-type IJJs, suggesting that $\Delta T_{\rm h}$ for sample S302 is $\sim$0.23~K for $P_{\rm dis}=$0.5~$\mu$W in our measurements at the lowest temperature. 
These estimations also support that our results cannot be explained only by the self-heating effects. 
\begin{figure}
\includegraphics[width=\linewidth]{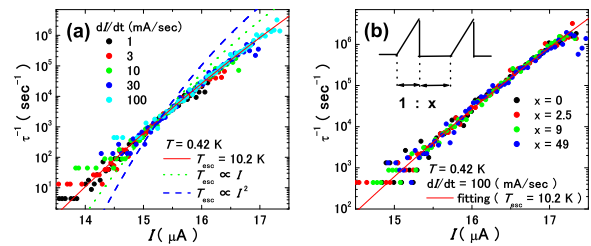}
\caption{(Color online). (a) $1/\tau(I)$ for the second switch of sample S302 at 0.42~K for different values of $dI/dt$. Corresponding pulse widths are 33.3~msec ($dI/dt$ = 1~mA/sec) $\sim$ 0.333~msec ($dI/dt$ = 100~mA/sec). (b) Similar plots of $1/\tau(I)$ for different waiting times between current pulses. Solid, dotted and dashed lines represent the fitting to Eq.(\ref{eq:tau_TA}) with the current-independent $T_{\rm esc}$(=10.2~K), with $T_{\rm esc} \propto I$ and with $T_{\rm esc} \propto I^2$, respectively.}
\label{fig:heating}
\end{figure}

In order to check the influence of the self-heating in a more quantitative manner, we measured $P(I)$ for the second switch by varying the values of $dI/dt$ and the waiting times between the current-ramp pulses, 
as shown in Figs.~\ref{fig:heating}(a) and \ref{fig:heating}(b), respectively. 
Note that the changes in $dI/dt$ correspond to those in the pulse widths since the amplitude of the ramped current pulse is fixed in the present measurement system.
We measured the bridge-type IJJs (sample S302) for this purpose since 
the effect of the self-heating process was expected to be more clearly observable. 
The experimental results indicated that $1/\tau(I)$ for the second switch was independent of the pulse widths and the waiting times. 
In terms of self-heating, this result can be understood if the experimental time scale is longer than a time scale of the crossover from the short-time limit to the long-time limit since the temperature rise is already saturated at the current of the second switch.\cite{Fenton}
However, even in such cases, $T_{\rm esc}$ should depend on the bias current, $I$, which determines the power density of the self-heating, giving rise to a convex curve in log-plots of $\tau^{-1}(I)$.
In other words, $T_{\rm esc}$ should be roughly proportional to $I^2$, or at least to a power of $I$ larger than one.
As shown in Fig.~\ref{fig:heating}(a), the measured data was compared with the calculated $\tau^{-1}(I)$, where $T_{\rm esc}$ was assumed to be constant (solid line), proportional to $I$ (dotted line), or proportional to $I^2$ (dashed line), while $I_c^{\rm 2nd}$ was fixed to the value listed in Table I.
We found that it was possible to successfully fit $\tau^{-1}(I)$  with the solid line rather than the dotted or the dashed lines over the whole range of the current distribution measurements. 
These results strongly suggest that the saturated behavior of $T_{\rm esc}$ below $T_{s}^{\rm 2nd}$ must be explained by considering the contribution of effects other than the self-heating effects in the resistive state after the first switch. 

Perhaps, this unusual behavior implies an increase of the electron temperature due to the injection of quasiparticles in the resistive state after the first switch, as previously discussed in terms of the modified heating theory of non-equilibrium superconductors. \cite{Parker}
In this theory, the injected quasiparticles are assumed to interact with phonons with an energy greater than 2$\Delta$, where $\Delta$ is the superconducting energy gap.
However, this mechanism seems to be unable to explain the increase in the electron temperature, which is as large as $\sim$10~\% of $T_c$, as discussed in the Appendix.
Furthermore, in the mesa-type IJJs with a few junctions, the second switch should occur at a junction close to that at which the first switch occurs.
On the other hand, in the case of the bridge-type IJJs, the junction at which the second switch occurs is not necessarily close to the junction at which the first switch occurs.
Therefore, the mechanism of quasiparticle injection is expected to be different between both types of IJJs.
Our results contradict this expectation.
\begin{figure}
\includegraphics[width=\linewidth]{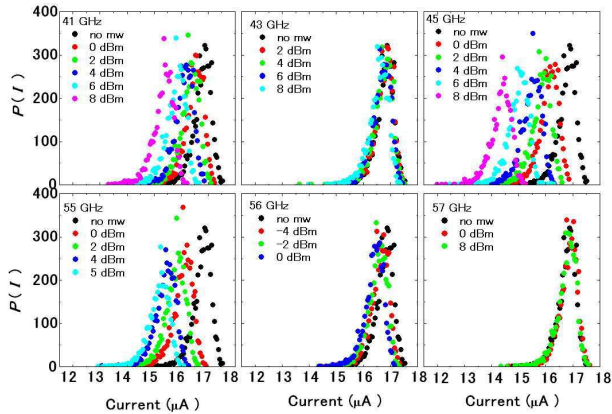}
\caption{(Color online). $P(I)$ for the second switch of sample S302 under microwave irradiation at several frequencies, which was measured at 0.42~K.}
\label{fig:mw2nd}
\end{figure}

Another possible candidate is the MQT process for the second switch. 
In this scenario, the disagreement between $T_{s}^{\rm 2nd}$ and $T_{\rm cr}^{\rm 2nd}$ should be resolved self-consistently. 
Note that $T_{\rm cr}^{\rm 2nd}$ is estimated on the basis of the single junction model described in Sec.~\ref{sec:model}, where the interlayer coupling effect is ignored. 
Thus, a more general treatment of $T_{\rm cr}$ in the multi-junction system is essential for resolving this issue. 

In order to examine the possibility for the occurrence of MQT for the second switch, we also attempted to observe a resonant peak in $P(I)$ for sample S302 by microwave irradiation. 
As shown in Fig.~\ref{fig:mw2nd}, we did not observe any resonant peaks in $P(I)$, although the distribution of $P(I)$ was slightly shifted to the lower current region due to the strong microwave irradiation. 
However, these results do not rule out the possibility of MQT for the second switch. If the crossover temperature for the second switch is given by $T_{s}^{\rm 2nd}$, the energy-level spacing $E_{\rm 01}$ seems to exceed 1~THz, 
suggesting that an $n$-photon process ($n>20$) is required in order to observe ELQ. 
It is almost impossible to observe such higher-ordered multi-photon processes in the present experimental setup. 
Rather, the possibility of ELQ for the second switch should be tested by applying direct THz radiation, which will be performed in the near future.

\section{\label{sec:conclusion}conclusion}
We studied the MQT behavior of IJJs in BSCCO single crystals by comparing a small mesa, which is a few nanometers thick and consists of two or three IJJs, with narrow bridge structure consisting of a few hundreds of IJJs. 
We carefully distinguished the switched junction from the remaining junctions in the repeated measurements in order to obtain the switching current distribution for both types of IJJs. 
For the first switching from the zero-voltage state, we observed the following three distinct regimes in the temperature dependence of $P(I)$ for both types of IJJs: (i) the phase-retrapping regime at high temperatures, (ii) the TA regime at intermediate temperatures, and (iii) the MQT regime at low temperatures. 
In the TA regime, the measured $P(I)$ for the mesa-type IJJs were in excellent agreement with the conventional theory of a single Josephson junction. 
Together with the observation of a multiphoton absorption process between the quantized energy levels at the lowest temperature, we confirmed that the observed MQT behaviors were intrinsic to a single IJJ in BSCCO and independent of the device structure. 

For the second switch (from the first resistive state), we found that the temperature-independent behavior of $T_{\rm esc}$, which is similar to the MQT behavior in the first switch, survived up to a much higher temperature than the crossover temperature estimated from the conventional single Josephson junction model. 
Based on our comparative study of two types of IJJs, we concluded that this unusual behavior for the second switch could not be explained in terms of the self-heating in the resistive state after the first switch. 
As possible candidates for the explanation of these observations, the MQT process for the second switch and the effective increase of electronic temperature due to quasiparticle injection were discussed. 
Unfortunately, the origin of this unusual behavior remains unknown at present. 
Although the MQT behavior for the first switch of IJJs in the BSCCO system is well established in this work, the understanding of the second switching phenomenon remains incomplete. 
The following two approaches are required in order to obtain an important key to resolving this issue. 
One is the full treatment of the MQT rate considering the interlayer coupling effects in the stacked IJJs. 
The other is the direct observation of a resonant peak in $P(I)$ under THz irradiation. 
The multi-dimensional aspects and the interlayer coupling effects are prominent features of IJJs systems in high-$T_c$ cuprates. 
We believe that the phase dynamics of this system will attract considerable interest in the exploration of new physics in multi-junction systems and the application of phase qubits to future quantum information processing. 

\section*{ACKNOWLEDGEMENTS}
We thank T. Koyama, S. Kawabata, S. Sato, and S. Kashiwaya for useful discussions and comments. This work has been supported by the PRESTO project of Japan Science and Technology Agency (JST) and 
the Grant-in-Aid for Scientific Research (19560011 and 18560011) 
from the Ministry of Education, Science, Sports and Culture of Japan. 
K. Ota also thanks the Japan Society for the Promotion of Science for financial support.
\appendix*
\section{}
In the modified heating theory proposed by Parker, \cite{Parker} it is assumed that the injected quasiparticles (QPs) interact with phonons with energy greater than $2\Delta$, where $\Delta$ is the superconducting energy gap. 
The energy distribution of non-equilibrium QPs is determined by that of phonons, which is described by a thermal distribution with an effective temperature $T^*$. 
Thus, the properties of non-equilibrium QPs and phonons are described by a set of rate equations proposed by Rothwarf and Taylor, \cite{Rothwarf} 
\begin{eqnarray}
\frac{dN}{dt}&=&I_0+\frac{2N_\omega}{\tau_B}-RN^2,\\
\frac{dN_\omega}{dt}&=&\frac{RN^2}{2}-\frac{N_\omega}{\tau_B}-\frac{N_\omega-N_{\omega T}}{\tau_\gamma},
\label{rate_eq}
\end{eqnarray}
where $N$ is the number of QPs, $N_\omega$ is the number of phonons with an energy greater than $2\Delta$, $I_0$ is the rate of externally injected QPs, $R$ is a recombination coefficient, $\tau_B$ is the characteristic time at which phonons create QPs by a pair-breaking process, and $\tau_\gamma$ is the characteristic time at which phonons disappear by processes other than the creation of QP. 
$N_{\omega T}$ is the number of thermal equilibrium phonons with an energy greater than $2\Delta$. 
The steady-state solutions of the rate equations are given as follows, 
\begin{eqnarray}
\Big(\frac{N}{N_T}\Big)^2&=&1+\Big(1+\frac{\tau_\gamma}{\tau_B}\Big)\frac{I_0\tau_R}{N_T},\label{eq:steady-state1}\\
\frac{N_\omega}{N_{\omega T}}&=&1+\frac{\tau_\gamma}{2N_{\omega T}}I_0,
\label{eq:steady-state2}
\end{eqnarray}
where $N_T(=\sqrt{2N_{\omega T}/R\tau_B})$ is the number of thermal equilibrium QPs, and $\tau_R(=1/RN_T)$ is the characteristic time at which QPs recombine into Cooper pairs. 

Eqs.~(\ref{eq:steady-state1}) and (\ref{eq:steady-state2}) suggest that $N_\omega/N_{\omega T}\sim(N/N_T)^2$ in the limit that $\tau_\gamma\gg\tau_B$, where the pair-breaking by phonons was more dominant than the other relaxation process of phonons. 
In the $T^*$-model by Parker, the effective temperature, $T^*$, which characterizes the distribution function of QPs through that of phonons, is related to the ratio $(N/N_T)$, 
\begin{equation}
\Big(\frac{T^*}{T}\Big)^3\sim\Big(\frac{N}{N_T}\Big)^2
\Big[\int^\infty_{X}\Big(\frac{x^2dx}{e^x-1}\Big)\Big/\int^\infty_{X^*}\Big(\frac{x^2dx}{e^x-1}\Big)\Big], 
\label{T*}
\end{equation}
where $X$ (or $X^*$) is $2\Delta(T)/k_BT$ (or $2\Delta(T^*)/k_BT^*$). 
Since the integral part included in Eq.~(\ref{T*}) is less than unity, 
$(N/N_T)^2$ gives the upper limit of the ratio $(T^*/T)^3$. 
In the limit of $\tau_r/\tau_B\gg 1$, the following relation is suggested, 
\begin{equation}
\Big(\frac{T^*}{T}\Big)^3<1+\frac{\tau_\gamma}{\tau_B}\frac{I_0\tau_R}{N_T}=1+\frac{\tau_\gamma I_0}{\tau_B N_T^2 R}. 
\end{equation}
Here, we used the relation $\tau_R=1/RN_T$.
The inverse of the recombination coefficient, 1/$R$, and the phonon pair-breaking time, $\tau_B$, for $d$-wave superconductors were microscopically estimated by J. Unterhinninghofen {\it et al}.\cite{Unterhinninghofen}
The estimated values of 1/$R$ and $\tau_B$ for $d$-wave superconductors were $\sim$10~ps and $\sim$1~fs, respectively.
A naive estimation of $\tau_\gamma$ using the thickness of the superconducting layer ($\sim$ 0.3 ~nm) and the velocity of sound ($\sim 10^3$~m/s) \cite{Sound velocity} suggests that $\tau_\gamma\sim 1\times 10^{-12}$ sec.
We also estimated the number of thermal equilibrium QP's, $N_T$, as $N_T\sim3\times 10^4$ by using the relation $N_T\sim\sigma_1 m^*/e^2\tau$ in the microwave conductivity, $\sigma_1\sim5\times10^6$~$\Omega^{-1}$m$^{-1}$[\onlinecite{Bonn}], $1/\tau\sim 1\times 10^{11}$per sec\cite{Shibauchi}, and the junction size $\sim 1~\mu$m$\times 1~\mu$m$\times$1.5~nm.
In our experiments, if we take the values of $I_0(=I_{\rm sw}^{\rm 2nd}/e)\sim 1\times 10^{14}$ per sec and $T$=0.4~K, the effective temperature is required to be $T^*<0.4+1\times10^{-4}$~K which is much smaller than the $T_{s}^{2nd}$.
Thus, it seems that our experimental result of high $T_{s}^{2nd}$ cannot be explained by the increase of the electron temperature due to the injection of quasiparticles in the resistive state after the first switch.

So far, we have discussed the pair-breaking process by phonons.
It is argued that the pairing mechanism of high-$T_c$ cuprates might be derived from the antiferromagnetic spin fluctuation rather than the electron-phonon interaction.
Even in that case, essential part of the above discussion is valid, independent of the mechanisms of the pair-breaking.


%
\newpage 

\begin{thebibliography}{70}
\bibitem{FIB} Regarding the S301 and S302 samples, we attempted to cut an IJJ stack into a trapezoidal shape in order to obtain IJJs with different critical currents at each junction. However, it was found that the angle between two slits shown in Fig.~\ref{fig:measurement}(b) was too small to obtain such junctions since the adjacent junctions were too close to each other on an atomic scale, while the lateral size is of the order of micrometers. Thus, the reduced switching currents for the first and the second switch in our samples are not related to the trapezoidal shape of the stack.
\bibitem{fit} In order to obtain the best fitting by using Eq.~(\ref{eq:tau_TA}), the least square fitting method was adopted for $log(\tau^{-1}(I))$ using both $I_c$ and $T_{\rm esc}$ as fitting parameters. 
We also used the same method for the second switch.
The values of $I_c$ and $I_c^{\rm 2nd}$, which were obtained at the lowest temperature, are listed in Table I. 
The temperature variation of $I_c$ and $I_c^{\rm 2nd}$ was less than 2\% for the measured samples, except for region (iii), as described in the text.
The temperature dependence of $T_{\rm esc}$ for the first and the second switch is shown in Fig.~\ref{fig:Tesc}.
\end{thebibliography}


\end{document}